\documentclass{emulateapj}

\usepackage{apjfonts}

\def\deg{$^{\circ} $}

\shorttitle{V405 Aur: A High Magnetic Field IP}
\shortauthors{Piirola et al.}

\begin{document}


\title{V405 Aurigae: A High Magnetic Field Intermediate Polar}


\author{V. Piirola\altaffilmark{1}, T. Vornanen, \& A. Berdyugin}
\affil{Tuorla Observatory, University of Turku, FI-21500 Piikki\"o,
Finland} \email{piirola@utu.fi}

\and

\author{G.V. Coyne, S.J.}
\affil{Vatican Observatory, V-00120 Citt\`a del Vaticano}


\altaffiltext{1}{Visiting Astronomer, Vatican Observatory, V-00120
Citt\`a del Vaticano}


\begin{abstract}
Our simultaneous multicolor ($UBVRI$) circular polarimetry has revealed nearly sinusoidal variation over the WD spin cycle, and almost symmetric positive and negative polarization excursions. Maximum amplitudes are observed in the $B$ and $V$ bands ($\pm$3\%). This is the first time that polarization peaking in the {\it blue} has been discovered in an IP, and suggests that V405 Aur is the highest magnetic field IP found so far. The polarized flux spectrum is similar to those found in
{\it polars} with magnetic fields in the range $B \sim$ 25-50 MG. Our low resolution circular spectropolarimetry has given evidence of transient features which can be fitted by cyclotron harmonics $n$= 6, 7, and 8, at a field of $B=31.5\pm0.8$ MG, consistent with the broad-band polarized flux spectrum. Timings of the circular polarization zero crossovers put strict upper limits on WD spin period changes and indicate that the WD in V405 Aur is currently accreting closely at the spin equilibrium rate, with very long synchronization timescales, $T_s > 10^9$ yr.
For the observed spin to orbital period ratio, $P_{spin}/P_{orb}$=0.0365, and $P_{orb} \sim$ 4.15 hr, existing numerical accretion models predict spin equilibrium condition with $B \sim$ 30 MG if the mass ratio of the binary components is $q_1 \sim$ 0.4. The high magnetic field makes V405 Aur a likely candidate as a progenitor of a polar.
\end{abstract}





\section{Introduction}

V405 Aur (RX J0558.0+5353) was discovered in the ROSAT all-sky survey (Haberl et al. 1994)
and was identified with a $V \sim$ 14.5 Cataclysmic Variable (CV), showing characteristics
of an Intermediate Polar (IP): the soft X-ray flux was modulated at 272.7 s,
which suggested that the system consists of a spinning magnetic white dwarf (WD)
accreting matter from a cool low-mass companion. An orbital period of 4.15 h was deduced
from optical spectroscopy. Further analysis of the ROSAT data
at energies higher than 0.7 keV revealed a period of 545 s, which is
the true spin period of the WD, and therefore the soft X-ray light curve has a
double-peaked structure (Allan et al. 1996; Evans \& Hellier 2004). Evidence of two-pole accretion has been found also from optical spectroscopy (Still et al. 1998; Harlaftis \& Horne 1999).

V405 Aur belongs to a small group of IPs which are unusually bright in soft X-rays,
and have also been found to emit polarized radiation in the optical and/or near IR, resembling
what we see in the strictly (or nearly) synchronous magnetic CVs, polars. The polarized IPs
known to date are BG CMi (Penning et al. 1986; West et al. 1987), PQ Gem (Piirola et al.
1993; Rosen et al. 1993; Potter et al. 1997), V2400 Oph (Buckley et al. 1997), V405 Aur
(Shakhovskoy \& Kolesnikov 1997), V2306 Cyg (Uslenghi et al. 2001; Norton et al. 2002),
and RX J2133.7+5107 (Katajainen et al. 2007). Of these IPs, PQ Gem was the first found
to show spin-modulated circular and linear polarization variations, arising from
cyclotron emission from two accretion regions near the opposite magnetic
poles of the spinning WD.

IPs are generally believed to have lower magnetic field ($B <$ 10 MG, see e.g.
Warner 1995) than polars, where the determined fields range from 7 MG
(V2301 Oph, Ferrario et al. 1995) to 230 MG (AR UMa, Schmidt et al. 1996). Typical
values for polars are a few tens of MG (see Cropper 1990 for a review), and the magnetic
field is sufficiently high to prevent the formation of an accretion disk. The
stream of matter from the companion is coupled to the magnetic field lines and
channeled via accretion columns onto the WD surface. In IPs accretion disk may exist,
but is truncated at some inner radius where flow to the WD takes place via accretion
curtains (Rosen et al., 1988; Ferrario et al., 1993).

IPs predominantly have longer orbital periods (4-5 h) than polars (typically $<$ 2 h),
though there is considerable overlap. Close binaries lose angular momentum via
gravitational radiation and magnetic wind braking, and therefore evolve towards shorter
orbital periods. For sufficiently high field strengths and small binary separations
the magnetic locking torque will start to spin-down the WD and synchronism may be reached,
the system thus becoming a polar. For IPs magnetic field measurements are difficult,
but from polarized flux spectrum estimates have been made for PG Gem (8-18 MG, Piirola
et al. 1993; 9-21 MG, V\"ath et al. 1996; Potter et al. 1997), V2400 Oph (9-20 MG,
V\"ath 1997), and RX J2133.7+5107 ($\gtrsim$ 20 MG, Katajainen et al. 2007). These
values are well within the range of the magnetic fields seen in polars and suggest that some of the
polarized IPs may evolve into polars. So far only one of the polarized IPs, PQ Gem, has
been shown to have the WD spinning down (Mason 1997).

In the present paper we report extensive multicolour ($UBVRI$) polarimetric observations
of V405 Aur, carried out in order to put constraints on the magnetic field
strength of the WD and on the system geometry. These observations were complemented with
low-resolution circular spectropolarimetry. We have also searched for possible spin
period changes by a long-term monitoring of phase shifts in the circular polarization
curves.

\section{Observations}


Broad-band multicolor polarimetry was carried out at the 2.5 m Nordic Optical Telescope
(NOT) at Roque de los Muchachos Observatory on La Palma. The instrument (TurPol) provides
strictly simultaneous measurements in five passbands close to the $UBVRI$ system, by
using four dichroic filters to split the light into the different spectral regions
(Piirola 1988). In the circular polarimetry mode the superachromatic quarter-wave
($\lambda/$4) plate was rotated in 90\deg\ steps above the plane-parallel calcite plate
polarizing beam splitter. For each orientation of the retarder, the two polarized
beams were integrated with a 25 Hz chopping frequency for a 5 s total integration
time each. With some dead time involved in the chopping procedure, the resulting time
resolution is about 12 s for photometry. One circular polarization observation
consists of two integrations with the $\lambda/$4 plate in orthogonal orientations,
and the time resolution is correspondingly about 24 s. In the simultaneous linear
and circular polarization mode the $\lambda/$4 plate was rotated in 22.5\deg\ steps.
One complete observation of linear and circular polarization consists of eight
integrations, and the time resolution is correspondingly about 1.6 minutes.
A summary of the polarization observations is given in Table 1.

\begin{deluxetable*}{rcccccr}
\tabletypesize{\scriptsize}
\tablecaption{Summary of polarimetric observations} \tablewidth{0pt} \tablehead{
\colhead{UT Date} & \colhead{HJD-24450000} & \colhead{Passbands} &
\colhead{Telescope} & \colhead{Instrument} & \colhead{Mode} &
\colhead{$N$}
} \startdata
1997 Nov 25 & 778.5566 - .6621&$UBVRI$&NOT&TurPol& SLC & 46\\
     Nov 27 & 780.5380 - .6806&$UBVRI$&NOT&TurPol& SLC &198\\
     Nov 28 & 781.5224 - .6103&$UBVRI$&NOT&TurPol& L &101\\
2001 Feb 04 &1945.4355 - .5224&$UBVRI$&NOT&TurPol&SLC & 36\\
     Feb 05 &1946.4423 - .5214&$UBVRI$&NOT&Turpol&SLC& 30\\
2002 Jan 25 &2300.3837 - .5019&$UBVRI$&NOT&Turpol& L & 96\\
2003 Jan 10 &2650.2704 - .3418&no filter&T-60&DIPOL& C &148\\
     Jan 13 &2653.3405 - .4482&  "  &T-60&DIPOL& C &136\\
     Jan 16 &2656.3313 - .4104&  "  &T-60&DIPOL& C &100\\
     Jan 30 &2670.3613 - .4923&  "  &T-60&DIPOL& C &150\\
     Jan 31 &2671.2289 - .3641&  "  &T-60&DIPOL& C &156\\
     Feb 07 &2678.2729 - .3721&  "  &T-60&DIPOL& C &116\\
     Feb 19 &2690.2659 - .3872&  "  &T-60&DIPOL& C &154\\
     Feb 20 &2691.2158 - .2945&  "  &T-60&DIPOL& C &100\\
     Mar 23 &2722.2757 - .3732&  "  &T-60&DIPOL& C &118\\
     Apr 27 &2757.3800 - .3950&  "  &NOT&ALFOSC& C & 48\\
     Sep 20 &2903.6210 - .6660&$UBVRI$&NOT&Turpol& SLC & 41\\
     Sep 21 &2904.6075 - .7285&$UBVRI$&NOT&Turpol& C &192\\
     Sep 22 &2905.6025 - .7385&$UBVRI$&NOT&Turpol& C &216\\
     Oct 18 &2931.5679 - .5967& $B$ &NOT&ALFOSC& L & 30\\
2004 Feb 26 &3062.3398 - .3597&no filter&NOT&ALFOSC& C & 64\\
     Feb 26 &3062.3713 - .4192&spectrop.&NOT&ALFOSC& C & 24\\
     Feb 28 &3064.4495 - .4805&no filter&NOT&ALFOSC& C & 96\\
     Feb 29 &3065.4080 - .4176&  "  &NOT&ALFOSC& C & 32\\
     Mar 01 &3066.3603 - .3921&  "  &NOT&ALFOSC& C & 96\\
 \enddata
\end{deluxetable*}

For WD spin period timing we made circular polarization observations of V405 Aur
also at the 60 cm reflector of Tuorla observatory, using a newly developed CCD
polarimeter DIPOL (Piirola et al. 2005). Because of the smaller aperture of the telescope,
the observations were done in unfiltered light. The polarimeter is equipped with
a thinned back illuminated CCD (Marconi) which has a high blue sensitivity. The time
resolution of these observations is about 1 min. Some complementary CCD-photometric observations
were carried out also at the 60 cm KVA telescope on La Palma.

Low resolution circular spectropolarimetry was done with the ALFOSC+FAPOL instrument
at the NOT. With the grism \#11 and 1.8" slit the spectral resolution was about 40{\AA}.
An achromatic $\lambda/$4 plate was rotated in 90\deg\ steps above the slit by the
polarimetry unit (FAPOL), controlled via the ALFOSC user interface. A plane parallel
calcite plate below the slit produced two perpendicularly polarized spectra on the
CCD. The two polarized beams passed through the collimator,
grism, and the camera onto the CCD. To improve time resolution, circular polarization
spectra were deduced from individual 120 s exposures. Null calibration was
done with the help of unpolarized standard stars. To check for the circular polarization
sign and scale, observations of the highly polarized star Grw+70 8247 were made.

\begin{figure}
\epsscale{1.12}
\plotone{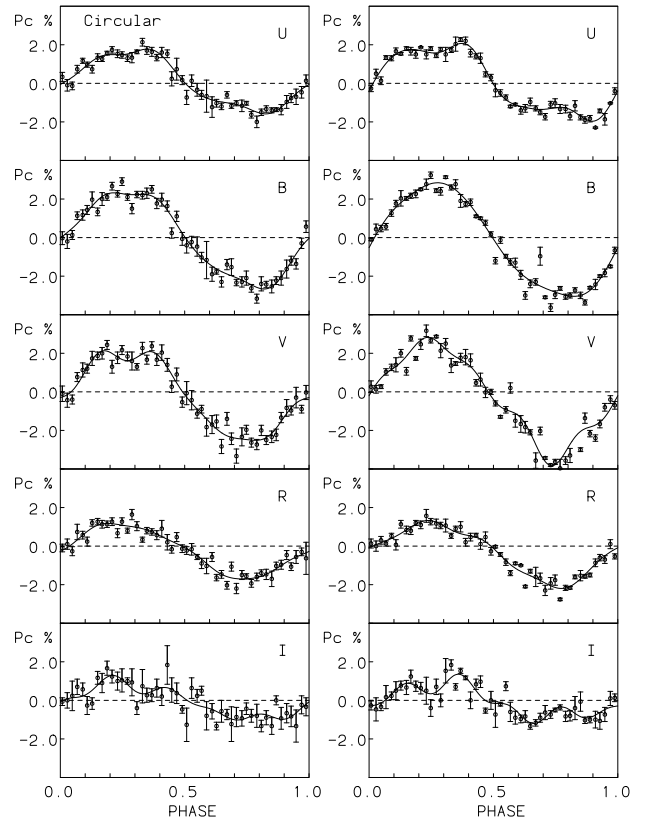} \caption{Simultaneous UBVRI circular
polarimetry of V405 Aur on 1997 Nov 27 (left) and 2003 Sep 22 (right).
Phases refer to the spin period of the WD given in Eq. 2.\label{fig1}}
\end{figure}

\begin{figure}
\epsscale{1.12}
\plotone{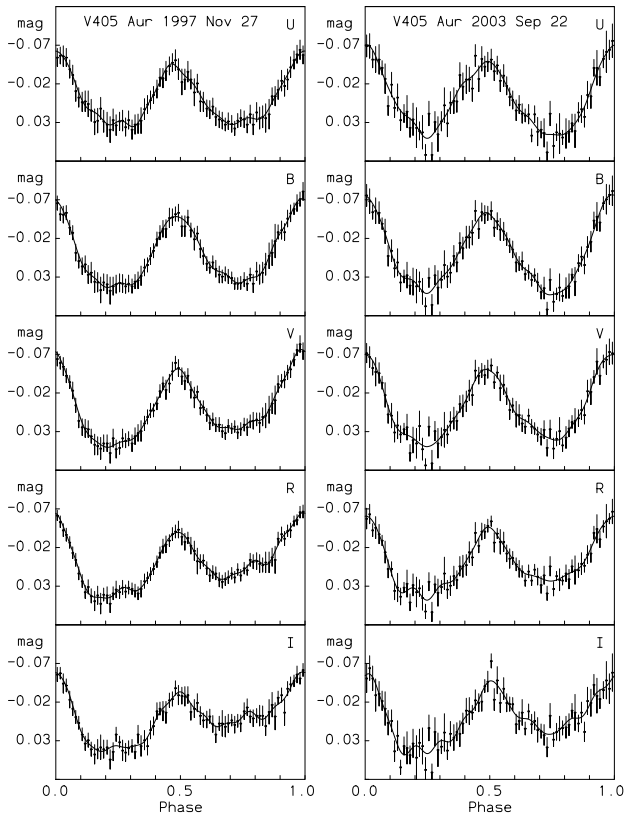} \caption{Simultaneous UBVRI
photometry of V405 Aur. Phases refer to the spin period of the WD given
in Eq. 2. \label{fig2_01}}
\end{figure}

\section{Results and discussion}

\subsection{Multicolor photopolarimetry}

Figure 1 illustrates our simultaneous $UBVRI$ circular polarimetry. The data collected
on two different nights show nearly sinusoidal and remarkably symmetric modulation,
with extreme values near $\pm$3\% in the $B$ and $V$, and about $\pm$2\% in the $U$ and
$R$ bands. In the $I$ band the modulation is smaller, about $\pm$1\%. This is the first
time that polarization peaking in the {\it blue} has been detected in an IP, and suggests
that V405 Aur is the highest field IP discovered so far. Its magnetic field is probably
similar to what is typically found in polars (25-50 MG). Nearly symmetric positive and
negative excursions in the circular polarization curves indicate that two opposite magnetic
poles (positive and negative) are accreting at similar rate in V405 Aur.

The left hand panels in Figure 1 show evidence of a small dip in circular polarization
near phase 0.25, where the positive magnetic pole points closest to us. This is due to
cyclotron beaming effect. Cyclotron emission intensity is at maximum in the direction perpendicular
to the magnetic field, and decreases to zero when looking along magnetic field lines.
Unpolarized background radiation from thermal sources dilutes the observed degree of
circular polarization when the polarized cyclotron flux decreases.

Our simultaneous $UBVRI$ light curves in Figure 2 show minima near the phases 0.25 and
0.75 where we are looking closest along the field lines, and maxima at the phases 0.0
and 0.5 where our line sight is perpendicular to the field lines, as seen from the
zero-crossings of circular polarization. This photometric behaviour is in accordance with
the cyclotron beaming effect, i.e., maximum emission takes place in directions
perpendicular to the magnetic field.

Many, but not all, polars show a pulse of linear polarization when our line of sight
is nearly perpendicular to the field lines. We have found no evidence of any linear
polarization pulse in V405 Aur. Linear polarization appears to be constant at the
level of $P_B \sim 0.36 \pm 0.03\% ,\ \theta \sim$ 138\deg $\pm$ 4\deg and is probably
of interstellar origin. Fitting the modified Serkowski law for the wavelength dependence of interstellar polarization (Whittet et al. 1992) to the average polarization in the
$UBVRI$ bands gives maximum polarization $P_{max}$ = 0.44 $\pm$ 0.04\%, at
$\lambda_{max}$ = 0.56 $\pm$ 0.11$\mu$m.

\begin{deluxetable*}{rrrccc}
\tabletypesize{\scriptsize}
\tablecaption{Spin cycle timings for V405 Aur} \tablewidth{0pt}
\tablehead{ \colhead{Cycle No.} & \colhead{HJD-24400000} &
\colhead{O-C(d)} & \colhead{$\sigma$(d)} & \colhead{Filter} &
\colhead{Note}
}
\startdata
      1. & 49681.47030 &  0.00012 & 0.00005 &no& 1\\
   1397. & 49690.28330 & -0.00003 & 0.00005 &no& 1\\
   2997. & 49700.38450 &  0.00013 & 0.00005 &no& 1\\
  13127. & 49764.33660 &  0.00004 & 0.00005 &no& 1\\
  14252. & 49771.43890 &  0.00005 & 0.00005 &no& 1\\
  14443. & 49772.64480 &  0.00014 & 0.00005 &no& 1\\
  16292. & 49784.31760 & -0.00007 & 0.00005 &no& 1\\
  16628. & 49786.43880 & -0.00009 & 0.00005 &no& 1\\
  16758. & 49787.25960 &  0.00000 & 0.00005 &no& 1\\
  16917. & 49788.26340 &  0.00001 & 0.00005 &no& 1\\
 173780. & 50778.56263 & -0.00004 & 0.00009 &U& 2\\
 173780. & 50778.56256 & -0.00011 & 0.00004 &B& 2\\
 173780. & 50778.56254 & -0.00013 & 0.00004 &V& 2\\
 173780. & 50778.56256 & -0.00011 & 0.00007 &R& 2\\
 173780. & 50778.56285 &  0.00018 & 0.00006 &U& 3\\
 173780. & 50778.56280 &  0.00013 & 0.00005 &B& 3\\
 173780. & 50778.56282 &  0.00015 & 0.00005 &V& 3\\
 173780. & 50778.56272 &  0.00005 & 0.00008 &R& 3\\
 174093. & 50780.53873 &  0.00005 & 0.00004 &U& 2\\
 174093. & 50780.53868 & -0.00001 & 0.00003 &B& 2\\
 174093. & 50780.53861 & -0.00007 & 0.00003 &V& 2\\
 174093. & 50780.53858 & -0.00011 & 0.00004 &R& 2\\
 174093. & 50780.53866 & -0.00003 & 0.00007 &U& 3\\
 174093. & 50780.53864 & -0.00005 & 0.00005 &B& 3\\
 174093. & 50780.53861 & -0.00007 & 0.00005 &V& 3\\
 174093. & 50780.53860 & -0.00009 & 0.00006 &R& 3\\
 358613. & 51945.44078 &  0.00009 & 0.00007 &B& 2\\
 358613. & 51945.44077 &  0.00008 & 0.00007 &V& 2\\
 358613. & 51945.44080 &  0.00011 & 0.00009 &R& 2\\
 358613. & 51945.44078 &  0.00009 & 0.00002 &U& 3\\
 358613. & 51945.44069 &  0.00000 & 0.00002 &B& 3\\
 358613. & 51945.44073 &  0.00004 & 0.00003 &V& 3\\
 358613. & 51945.44072 &  0.00003 & 0.00003 &R& 3\\
 358772. & 51946.44450 &  0.00002 & 0.00010 &B& 2\\
 358772. & 51946.44458 &  0.00010 & 0.00014 &V& 2\\
 414836. & 52300.38478 & -0.00002 & 0.00003 &U& 3\\
 414836. & 52300.38479 & -0.00001 & 0.00004 &B& 3\\
 414836. & 52300.38474 & -0.00006 & 0.00002 &V& 3\\
 414836. & 52300.38476 & -0.00003 & 0.00002 &R& 3\\
 470258. & 52650.27201 & -0.00006 & 0.00005 &no& 4\\
 470745. & 52653.34659 &  0.00002 & 0.00007 &no& 4\\
 471218. & 52656.33269 &  0.00000 & 0.00004 &no& 4\\
 473441. & 52670.36694 &  0.00013 & 0.00007 &no& 4\\
 473578. & 52671.23165 & -0.00007 & 0.00003 &no& 4\\
 474694. & 52678.27720 &  0.00001 & 0.00009 &no& 4\\
 476594. & 52690.27218 &  0.00001 & 0.00004 &no& 4\\
 476744. & 52691.21913 & -0.00001 & 0.00002 &no& 4\\
 481664. & 52722.27984 &  0.00001 & 0.00004 &no& 4\\
 487224. & 52757.38082 & -0.00011 & 0.00002 &no& 5\\
 510545. & 52904.60991 &  0.00006 & 0.00003 &U& 2\\
 510545. & 52904.60991 &  0.00006 & 0.00002 &B& 2\\
 510545. & 52904.60989 &  0.00004 & 0.00003 &V& 2\\
 510545. & 52904.60985 &  0.00000 & 0.00004 &U& 3\\
 510545. & 52904.60973 & -0.00012 & 0.00003 &B& 3\\
 510545. & 52904.60979 & -0.00006 & 0.00002 &V& 3\\
 510545. & 52904.60975 & -0.00010 & 0.00003 &R& 3\\
 510545. & 52904.60979 & -0.00006 & 0.00003 &I& 3\\
 510703. & 52905.60741 &  0.00009 & 0.00002 &U& 2\\
 510703. & 52905.60736 &  0.00004 & 0.00002 &B& 2\\
 510703. & 52905.60731 & -0.00001 & 0.00002 &V& 2\\
 510703. & 52905.60728 & -0.00004 & 0.00003 &U& 3\\
 510703. & 52905.60735 &  0.00003 & 0.00002 &B& 3\\
 510703. & 52905.60734 &  0.00002 & 0.00002 &V& 3\\
 510703. & 52905.60734 &  0.00002 & 0.00002 &R& 3\\
 510703. & 52905.60733 &  0.00001 & 0.00002 &I& 3\\
 535530. & 53062.34390 &  0.00006 & 0.00005 &no& 5\\
 535530. & 53062.34373 & -0.00011 & 0.00004 &no& 6\\
 535864. & 53064.45240 & -0.00003 & 0.00003 &no& 5\\
 536016. & 53065.41211 &  0.00008 & 0.00005 &no& 5\\
 536016. & 53065.41195 & -0.00008 & 0.00005 &no& 6\\
 536167. & 53066.36538 &  0.00006 & 0.00002 &no& 5\\
 702502. & 54116.46291 &  0.00018 & 0.00008 &no& 7\\
\enddata
\tablecomments{(1) Photometry from Allan et al. 1996, (2) $UBVRI$
polarimetry at the NOT, (3) $UBVRI$ photometry at the NOT (4) CCD polarimetry at T-60, (5) CCD
polarimetry at the NOT, (6) CCD photometry at the NOT,
(7) CCD photometry at the KVA-60}
\end{deluxetable*}

The zero-crossings of circular polarization give an accurate measure of the rotation
of the WD. We have used these timings to search for possible changes in the WD spin period
and evidence of spin-up or spin down. Table 2 lists our observations of the time
of the positive cross-over of circular polarization. We have included in Table 2 also a few
photometric timings to extend the time span. Maxima in light curves take place at the circular polarization zero-crossings, and our ephemeris from circular polarization is accurate enough to solve without ambiguity which of the two maxima corresponds to the positive cross-over of circular polarization. Weighted second order polynomial
fit to the timings obtained over about 12 year interval yields a greatly improved
ephemeris for the WD spin:
\begin{eqnarray}
T_\circ= HJD \ 2\ 449\ 681.46389(5)\nonumber \\ + \ 0.0063131474(4)\hfill \times E + \ 4(4)\times 10^{-16}\times E^2
\end{eqnarray}

The numbers in parentheses give the error estimates in units of the last digit of
the determined coefficient. No statistically significant second-order term is detected.
A weighted linear fit to
the timings gives the following ephemeris for the positive crossover of circular
polarization:
\begin{eqnarray}
T_\circ= HJD \ 2\ 449\ 681.46387(2)\nonumber \\ + \ 0.0063131476(3)\times E.
\end{eqnarray}

The residuals (O-C) of the observed times of positive crossover from those predicted by the
linear ephemeris are shown in Figure 3. There is no clear trend suggesting any significant
higher-order terms. The rather tight limits for WD spin period changes imply very long
spin-down timescales ($T_s > 10^9$ yr) for V405 Aur in its current evolutionary and accretion state.

\begin{figure}
\epsscale{1.1}
\plotone{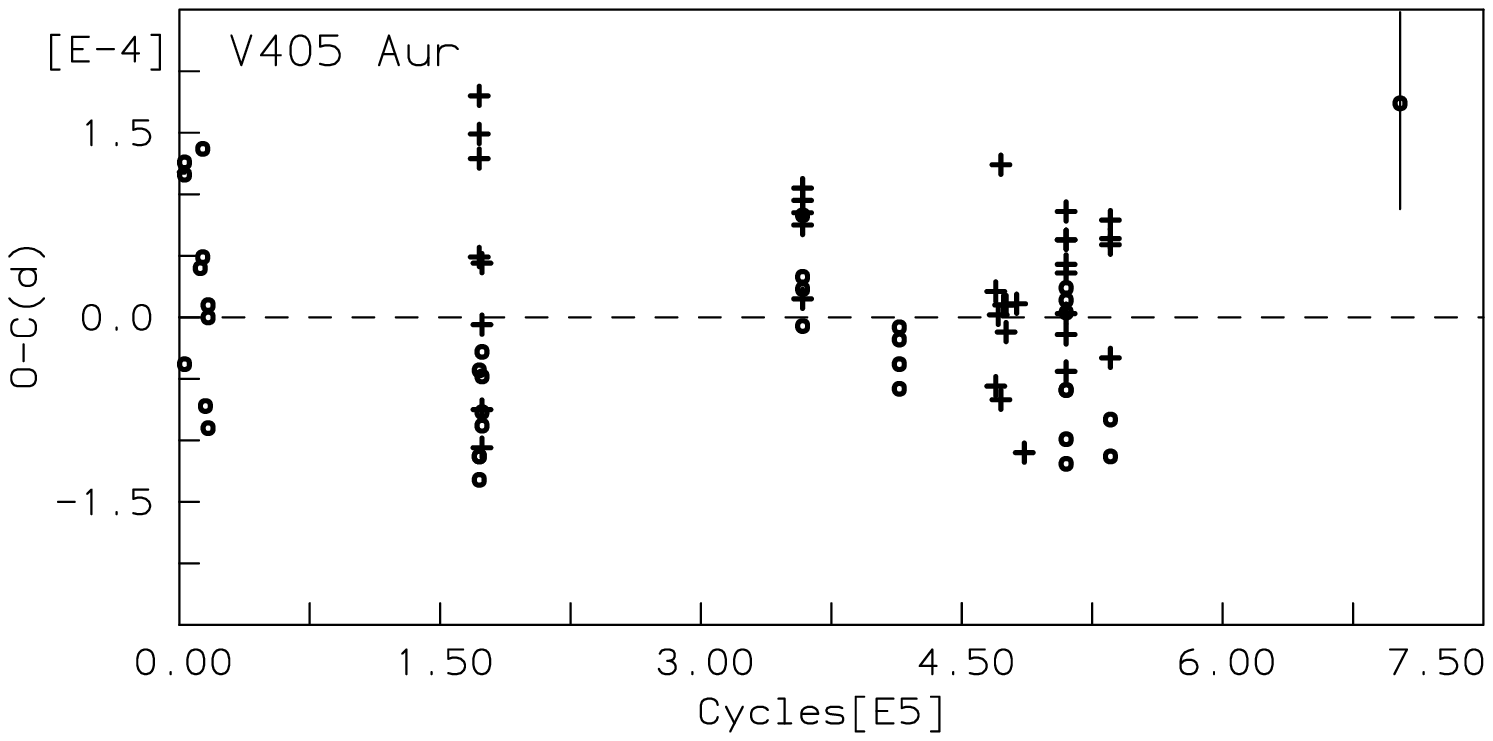} \caption{ Spin cycle timings of V405 Aur.
Residuals from the weighted linear fit (Eq. 2) are plotted vs. cycle
number. Different symbols denote photometry ($\circ$) and polarimetry (+).\label{fig2_02}}
\end{figure}

\subsection{Cyclotron model fittings}

Observed polarized {\it fluxes} are independent of any unpolarized background
sources (accretion stream, disk, WD photosphere) and therefore very useful for comparing with
existing cyclotron models. Broad-band polarized flux spectrum may give rough estimates
of the magnetic field in the cyclotron emission region, as cyclotron emission intensity and polarization are characteristic to each harmonic $n = \omega/\omega_c$  of the fundamental cyclotron frequency $\omega_c$ (see e.g. Wickramasinghe \& Meggitt 1985). The wavelength $\lambda_c$ which corresponds to $\omega_c$ is in turn related to the magnetic field strength by $\lambda_c \sim 108\mu$m/$B$[MG].
The method has been applied to the polarized IPs PQ Gem (Piirola et al. 1993; V\"ath et al. 1996; Potter et al. 1997) and V2400 Oph (V\"ath 1997) and the suggested field strengths are in the range 8-20 MG in both of these objects.

Polarized flux spectrum is strongly dependent on the parameters of the adopted cyclotron model and the system geometry and this limits the usefulness of the method. However, it may be the only way of getting any magnetic field estimates at all for IPs which do not show measurable cyclotron harmonic features or photospheric Zeeman features in their spectra, because of the overlaying strong disk and stream emission.

Figure 4 gives the peak circularly polarized fluxes in the $UBVRI$ bands, computed from the
data displayed in Figures 1 and 2, compared with those deduced from existing cyclotron
models. With constant temperature $\Lambda = 10^6$ models (Wickramasinghe \& Meggitt 1985) the polarized flux distribution is too narrow (dashed and dotted lines). Broader distribution and better fit is obtained for extended ribbon-like accretion shocks model of Wickramasinghe, Wu, \& Ferrario (1991), where allowance is made for field spread and for the change in shock height as a function of
specific accretion rate. The fit shown in Figure 4 (solid line) corresponds to a field of
$B \sim$ 26 MG.

\begin{figure}
\epsscale{1.1}
\plotone{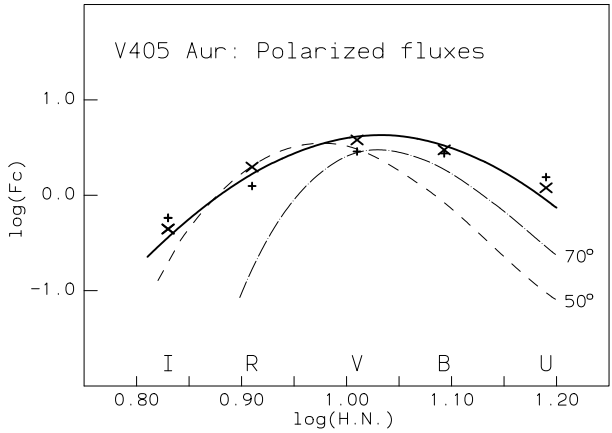} \caption{Observed peak circularly polarized
fluxes of V405 Aur, compared with those deduced from existing cyclotron models. The solid line corresponds to the extended emission region model (Wickramasinghe et al. 1991) with a field of 26 MG and average view angle $\sim$60\deg. Dot-dashed and dashed lines give fluxes from the 20 kEV constant temperature, $\Lambda = 10^6$, model (Wickramasinghe \& Meggitt 1985) for viewing angles $\alpha$ = 70\deg\ and 50\deg.}
\end{figure}

Without measurable linear polarization from the cyclotron source it is not possible
to put strict constraints on the geometric model, which is defined basically by the inclination, $i$, of the WD spin axis and the colatitude, $\beta$, of the cyclotron emission region(s). This
adds further uncertainty in the modeling of the cyclotron flux spectrum (Figure 4), which is
dependent on the angle between our line of sight and the magnetic field. The symmetric
shape of the circular polarization curves over the WD spin period (Figure 1) requires that $\beta$ is large. This was noted already by Shakhovskoy \& Kolesnikov (1997), who suggested a nearly equatorial magnetic field in V405 Aur.

The small dip at the circular polarization maximum (Figure 1, left panel) indicates
that when the emission region is pointing closest to us the viewing angle, $\alpha$, reaches
a value where the beaming effect starts to reduce cyclotron flux, but does not yet drop it
substantially. Less clear evidence of a beaming effect is seen on the right hand panel of Figure 1 suggesting that probably the emission region had moved slightly towards larger $\alpha$ values, and consequently the observed cyclotron emission was less influenced by the beaming effect on that night.

With the above constraints from circular polarization curves, we have tried to reproduce the observed polarization and light curves of V405 Aur with a geometric model involving extended emission strips on the WD surface. We have applied the viewing angle dependence of the cyclotron flux polarization and intensity from Wickramasinghe \& Meggitt (1985), as done in the earlier work by Piirola et al. (1990, 1993, 1994), where the models are also explained in more detail. The minimum total flux determined from each of the $UBVRI$ light curves was adopted as the wavelength dependent unpolarized background flux in the model computations. This is a good approximation as the relative amount of cyclotron emission in the total observed flux is small in V405 Aur. Centered, or slightly de-centered, dipole field was assumed for computing the direction of field lines relative to the normal to the WD surface for the emission arcs located off from the magnetic poles.

The nearly symmetric rising and descending parts of the circular polarization curves suggest that the tilt of magnetic field lines with respect to the normal to the WD in the emission region is not large in the direction of longitude, i.e., the cyclotron region is not located far from the magnetic pole in that direction. Such accretion geometry may be favored in spin equilibrium case. In contrast, for PQ Gem Potter et al. (1997) have found the cyclotron arc located {\it ahead} of the magnetic pole in longitude, which is consistent with the {\it spin-down} of the WD in this system. Estimates from X-ray spectral fits (Evans \& Hellier 2004) give rather small values of the blackbody emitting area ($<$ 0.001 of the WD surface) associated with the accretion region. Hence, we assume relatively narrow cyclotron emission arcs in our model.

\begin{figure}
\epsscale{1.17}
\plotone{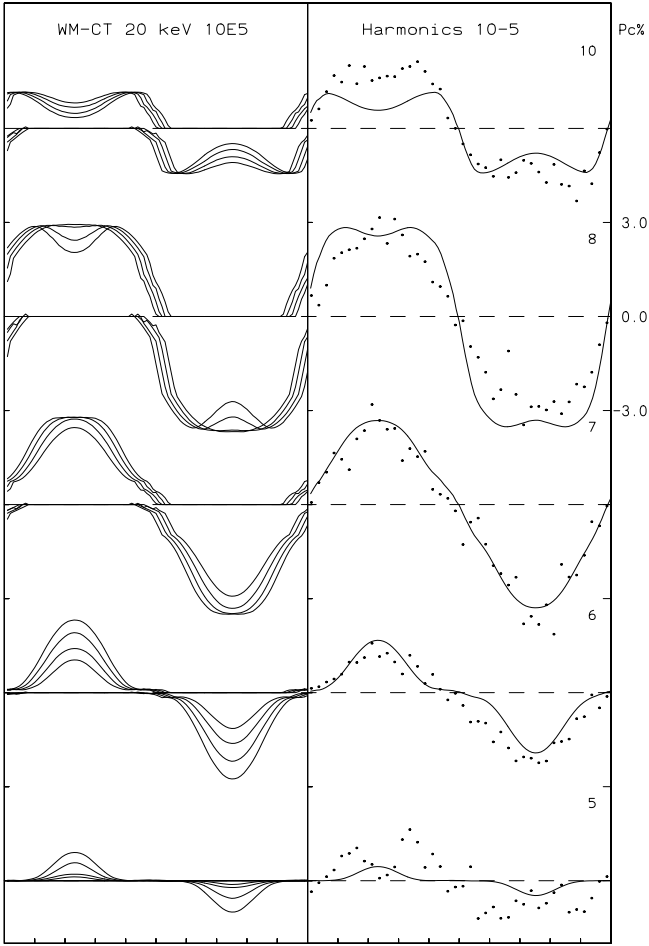}
\caption{Examples of simulated circular polarization curves for $i$ = 38\deg\ and two nearly
equatorial cyclotron emission regions extended by $\Delta\lambda$ = 24\deg\ in longitude and
confined between colatitudes $\beta_1$ = 80-92\deg and $\beta_2$ = 76-88\deg\ for the positive and negative regions, respectively. Each of the emission regions has been divided into four equidistant strips with 4\deg\ intervals in $\beta$ for the modeling shown in the left panel. The right hand panel displays curves from the accretion model shown in Figs. 7 and 8, compared with the circular polarization normal points of 2003 Sep 22 in the $UBVRI$ bands (from top to bottom).}
\end{figure}

\begin{figure}
\epsscale{1.19}
\plotone{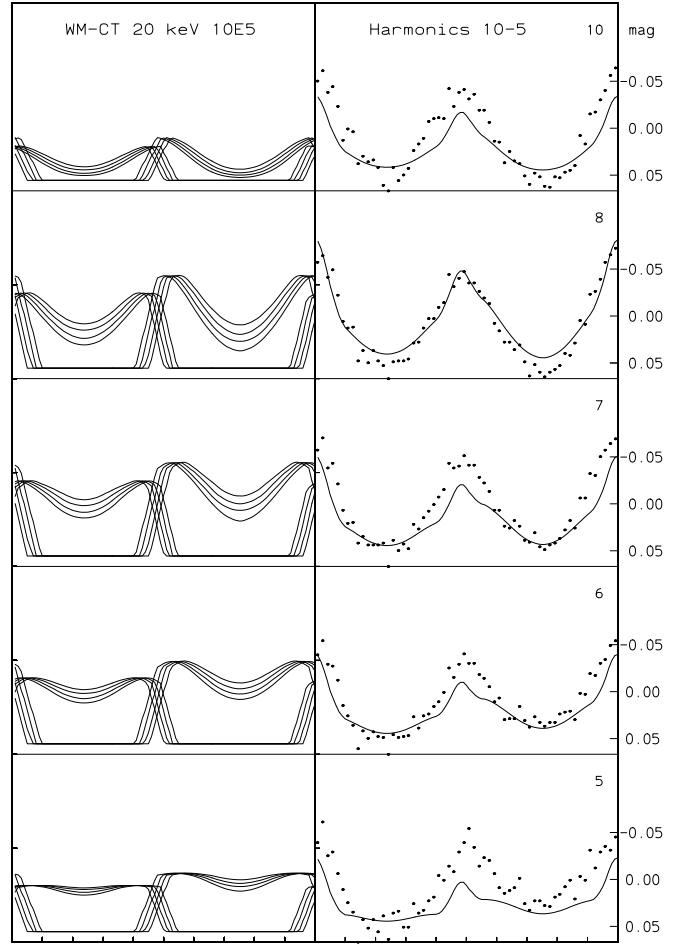}
\caption{Simulated light curves for the same models as in Figure 5, compared with the light curves from 2003 Sep 22 in the $UBVRI$ bands (dots, from top to bottom).}
\end{figure}

The main features of the circular polarization and light curves can be reproduced reasonably well (Figures 5 and 6), but the cyclotron flux from the constant temperature models drops too much when moving away from the wavelength of the peak polarized flux. Hence, the amplitude of the circular polarization and intensity variations becomes too small for the $U$ and $I$ bands (top and bottom panels, respectively). Also the beaming effect, the dip of polarization at phases 0.25 and 0.75, becomes too pronounced at high harmonics. Producing a broader cyclotron flux spectrum would require inhomogeneous emission regions with spread in the physical parameters, the electron density, temperature, and the magnetic field. Such modeling is, however, beyond the scope of the present paper.

The results from our computations shown in Figures 5 and 6 support the earlier suggestion by Shakhovskoy \& Kolesnikov (1997) that the magnetic dipole axis is nearly equatorial in V405 Aur (Figures 7 and 8). With such a large value of the colatitude, $\beta \sim$ 90\deg\ , best fits to our circular polarization and light curves are obtained with the WD spin axis inclination in the range $i \sim$ 30-50\deg. To put better constraints on the system geometry of V405 Aur, further efforts to establish possible variability of the linear polarization ($P, \theta$) would be required.

\begin{figure*}
\epsscale{0.75}
\plotone{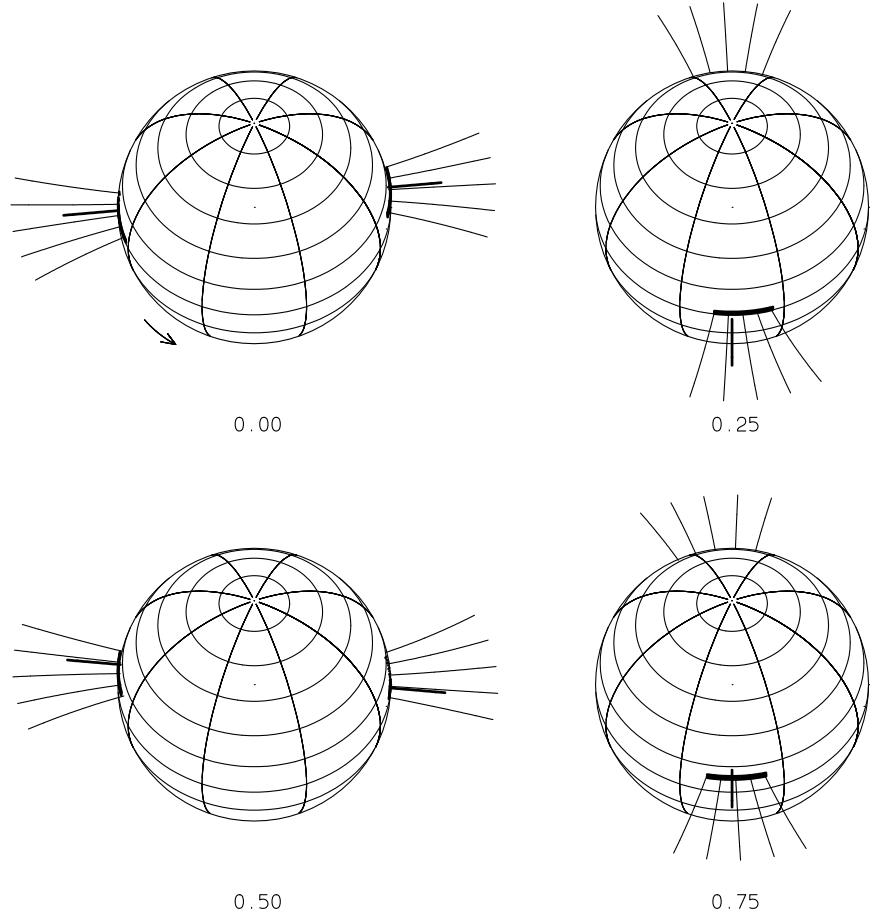} \caption{ Illustration of the model used to compute the circular
polarization and light curves shown in Figs. 5 and 6\label{fig2}. The spinning WD is seen at inclination $i$=38\deg\ and the magnetic dipole angle is $\beta$=82\deg . Positive magnetic pole points closest to us at the phase $\Phi\sim$0.25 and negative pole at $\Phi\sim$0.75. Brighter light curve maximum takes place at $\Phi \sim$ 0.0 and another maximum at $\Phi \sim$ 0.5. Cyclotron emission arcs and the dipole axis are drawn with thick lines. The emission arcs are off from the magnetic poles by $\Delta\beta_1$=-3\deg\ and $\Delta\beta_2$=+4\deg\ for the positive and negative pole, respectively. The centers of the 24\deg\ long emission arcs are ahead of the poles in longitude by $\Delta\lambda_1$=5\deg\ and $\Delta\lambda_2$=1\deg\ . The accretion model geometry is shown in Fig. 8.}
\end{figure*}

\begin{figure*}
\epsscale{1.1}
\plotone{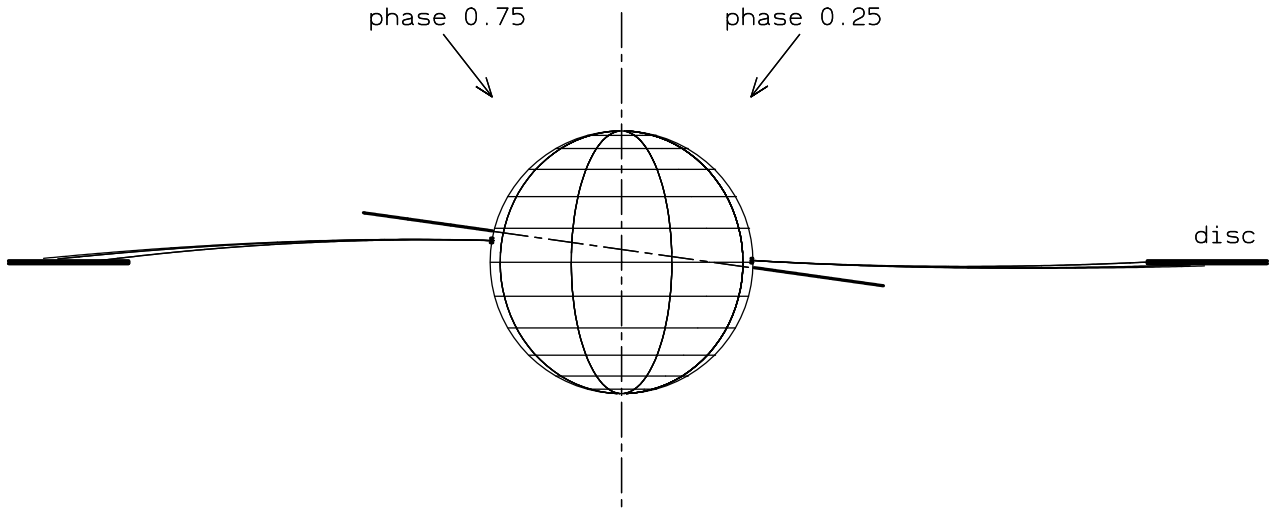}
\caption{Side-on schematic view of the accretion model, shown for $i$=38\deg\ and magnetic dipole angle $\beta$=82\deg. The dipole is decentered by 0.1$R_{WD}$ towards the upper rotational pole. Arrows give viewing directions at phases 0.25 and 0.75. For a larger inner radius of the disc the accreting field lines hit the WD surface closer to the magnetic poles.}
\end{figure*}

\subsection{Comparison with X-ray data}

Soft X-rays in polarized IPs are dominated by blackbody emission from the heated WD surface near the accretion shock. The double-peaked structure of the soft X-ray light curve in V405 Aur indicates that both accretion regions contribute. The equality of the two maxima requires that the angle between the magnetic and spin axes is high (Evans \& Hellier 2004), as also suggested earlier by the circular polarization curve (Shakhovskoy \& Kolesnikov 1997) and our modeling in the present paper (Sect 3.2).

The details of the geometric model of Evans \& Hellier (2004) from X-ray data are not fully supported by the circular polarization curves. With their values of spin and magnetic axis inclination ($i$=65\deg\ and $\beta_1$=60\deg) the upper magnetic pole passes near the center of the visible disk of the WD when pointing closest to us, and the corresponding viewing angles, $\alpha \sim$ 0\deg\, would give prominent cyclotron beaming effects: circular polarization would drop to zero in all wavebands. Our circular polarization curves (Figure 1) show only minor depressions at phases 0.25 and 0.75 in the $U$ band and marginal evidence in the $B$ and $V$ bands, suggesting that the viewing angle, $\alpha >$ 40\deg\ throughout the spin cycle. The symmetric polarization and light curves require the two accretion regions be nearly at the {\it same} latitude. For nearly centered dipole this means nearly equatorial magnetic field. More exotic possible configurations might be a grossly de-centered (in latitude) dipole, or quadrupole field with one positive and one negative pole visible (for parts of the spin cycle) and nearly at the same latitude.

Our improved ephemeris (Eq. 2) allows us to phase some of the published X-ray curves with our ($UBVRI$) circular polarization and light curves. In hard X-rays V405 Aur shows a {\it single} peak, and the epoch of the maximum given by de Martino et al. (2004), HJD$_{max}$ = {2 450 364.159695}, occurs at the phase 0.74 of our ephemeris. This is very near the phase (0.75) where the negative magnetic pole is pointing closest to us, and means that this pole is seen brighter in hard X-rays. Evans \& Hellier (2004) define phase zero of their XMM-Newton data as the peak of the larger of the soft X-ray maxima, which corresponds to HJD {2 452 187.55904}. With our ephemeris this takes place at the phase 0.44, which is near the second maximum in the optical light curves (phase 0.5), where we look nearly perpendicular to the magnetic field. This would be in accordance with the standard accretion curtain model (see e.g. Hellier et al. 1991), where the opacity of the column causes X-rays to emerge preferentially perpendicular to the curtain, and argues against the other plausible explanation that foreshortening of the accretion polecaps when viewed face on and near the limb of the WD would be the dominating effect on the soft X-ray flux observed in V405 Aur.
However, neither de Martino et al. (2004) nor Evans \& Hellier (2004) found the absorption dip expected from the accretion-curtain dip model, and the soft X-ray behaviour of V405 Aur vs. the magnetic phase requires further attention.

\subsection{Spectropolarimetry}

If cyclotron harmonics are detected in the spectrum, they provide an accurate method for determining
the magnetic field strength in the emission region. However, thermal broadening effects and magnetic field spread in an inhomogeneous region may smear out the harmonics and make them undetectable. Also the overlying emission from other than the cyclotron source need to be carefully modeled and removed (see e.g. Cropper et al. 1988). Because of these complications, cyclotron harmonic humps have not been seen in many of the known polars, and even if detected are visible only at some phase angle intervals during the WD spin cycle.

In V405 Aur the bright thermal emission from the disk and the accretion stream dilute the
cyclotron flux at least by an order of magnitude when compared with strongly polarized polars which have $P_c \sim$ 20-40\%. This makes the detection of cyclotron humps in the intensity spectrum of V405 Aur unlikely. Therefore, we have carried out circular spectropolarimetry with the aim to search for cyclotron harmonics in the {\it polarized flux} spectra, where the effects from thermal emission sources are largely eliminated.

Most of the polarized spectra we obtained show no clear cyclotron harmonic pattern.
However, at some phase angles of the WD spin period we see transient structures which suggest a possible cyclotron origin.  A single 120 s exposure of the spectrum of the negative emission region in Figure 9 shows three peaks (downwards), the inverse wavelengths of which can be fitted with cyclotron harmonics $n$ = 6, 7, and 8, at magnetic field of $B$ = 31.5$\pm$0.8 MG. If confirmed, this would make V405 Aur the first IP with a direct measurement of magnetic field, and the value found is typical of that of polars.

\begin{figure}
\epsscale{1.15}
\plotone{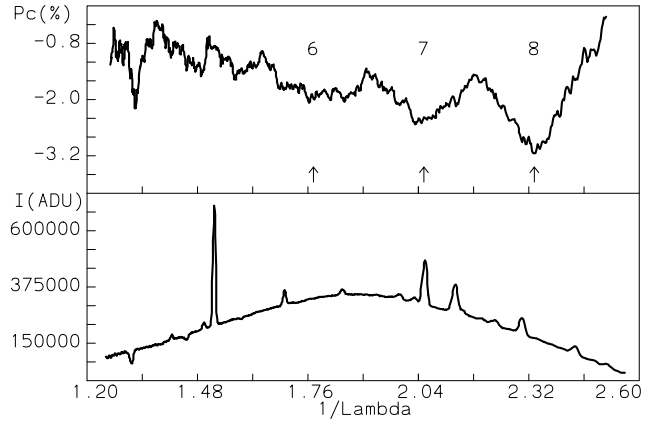}
\caption{Circular spectropolarimetry of V405 Aur on 2004 Feb 26.
The negative circular polarization spectrum (120 s exposure, top panel) shows features which can be fitted on the 1/$\lambda$ scale with cyclotron harmonics $n$ = 6, 7, and 8 from a field of $B$ = 31.5$\pm$0.8 MG.  The degree of circular polarization has been corrected for the diluting unpolarized flux from the emission lines and smoothed with a $\Delta\lambda \sim$ 70 {\AA} wide window filtering. The bottom panel gives a raw intensity spectrum (ADU). }
\end{figure}

\begin{figure*}
\epsscale{1.}
\plotone{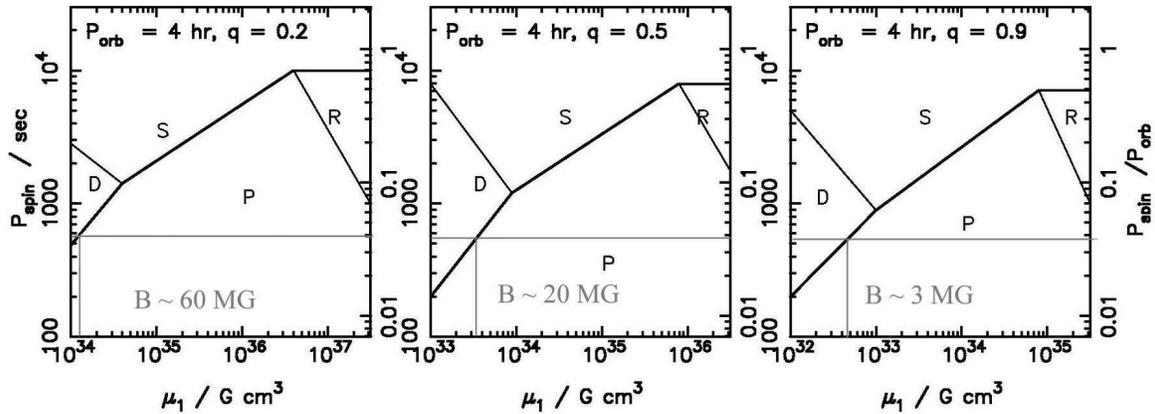} \caption{ The location of V405 Aur in the $P_{spin}/P_{orb}$ vs. magnetic moment $\mu_1$ diagram from Norton et al. (2008) for $P_{orb}$ = 4 hr and mass ratios $q$ = 0.2, 0.5, and 0.9. Spin equilibrium condition predicted by the numerical model is defined in these graphs by the boundary line between the D (disk) and P (propeller), or the S (stream) and P (propeller) type accretion. The respective surface magnetic field values (MG) correspond to a solar-mass WD.  \label{fig2_03}}
\end{figure*}

\subsection{Is V405 Aur a progenitor of a polar?}

The broad-band ($UBVRI$) circular polarization wavelength dependence (Fig. 1) is similar to what is seen in moderately high-field polars ($B \sim$ 25-50MG), such as VV Pup (Piirola et al. 1990; Schwope \& Beuermann 1997) and V834 Cen (Piirola 1995; Ferrario et al. 1992). Our magnetic field estimates for V405 Aur from the polarized flux spectrum ($B \sim$ 25 MG, Sect. 3.2) and spectropolarimetry ($B \sim $30 MG, Sect. 3.4) are in accordance with the values determined for those polars from  cyclotron humps or Zeeman spectroscopy. This makes V405 Aur a likely candidate as a progenitor of a polar.

Using a numerical model of magnetic accretion, Norton et al. (2004) have predicted that IPs with magnetic moments $\mu_1 \ge 5 \times 10^{33}$ G cm$^3$ and $P_{orb} >$ 3 hr will evolve into polars. For a solar-mass WD a surface field of $B$ = 30 MG corresponds to $\mu_1 \sim 5 \times 10^{33}$ G cm$^3$, which gives further support that V405 Aur may evolve into a polar.

The ratio of the spin to the orbital period $P_{spin}/P_{orb} = 0.0365$ indicates that V405 Aur is still far from synchronism. Our spin period analysis (Fig. 3 and Eqs. 1-2) gives $P_{spin}$ which is constant within tight limits, and the corresponding spin-down timescales are very long ($T_s > 10^9$ yr). Therefore, V405 Aur is currently accreting closely at the spin equilibrium rate. For small $P_{spin}/P_{orb}$ values this condition occurs in disk-like accretion (Norton et al. 2004, 2008). The angular momentum gain from accretion onto the WD balances the magnetic braking torque, and equilibrium is reached.

In Figure 10 we show the location of V405 Aur in the $P_{spin}$ vs. magnetic moment, $\mu_1$, diagram from Norton et al. (2008) for $P_{orb}$ = 4 hr, and three different mass ratios, q=0.2, 0.5, and 0.9. The inferred magnetic field strength depends strongly on the value of $q$. For the value of $P_{spin}/P_{orb}$ = 0.0365 these numerical model computations predict a field $B$ = 20 MG for $q$ = 0.5 and the spin equilibrium condition, defined in these graphs by the boundary line between the D (disk) and P (propeller), or the S (stream) and P (propeller) type accretion. Lower $q$ implies larger $B$ (60 MG for $q$ = 0.2). $B$=30 MG would be obtained for $q \sim$ 0.4 from these models, assuming a solar mass WD in the conversion of $\mu_1$ to the surface field.

Because of the angular momentum loss via gravitational radiation and magnetic wind braking, V405 Aur will evolve towards shorter orbital period and smaller binary separation, $a$. It is expected that when small enough $a$ is reached, the magnetic braking torque
will overcome the spin-up accretion torque and the WD will start to spin-down towards synchronization, the system thereby becoming a polar.

\section{Conclusions}

Our simultaneous multicolor ($UBVRI$) circular polarimetry of the intermediate polar V405 Aur has revealed polarization which peaks in the $B$ and $V$ passbands, similar to relatively high field ($B=25-50$ MG) polars. This is the first time that a polarized flux spectrum reaching maximum at such short wavelengths has been found in an IP, and suggests that V405 Aur is the highest field IP found so far. We have also detected transient features in circularly polarized spectra which can be fitted
on the wavenumber scale with cyclotron harmonics $n$=6, 7, and 8, at a magnetic field of $B=31.5\pm0.8$ MG. Such a field is consistent with the broad-band polarized flux spectrum observed, and this makes V405 Aur a likely candidate of a progenitor of a polar.

Nearly sinusoidal and symmetric positive and negative circular polarization excursions in all of the $UBVRI$ passbands indicate that the two opposite magnetic poles are accreting at very similar rate, and are located far from the rotational poles, probably near the equator. For such a magnetic geometry, our numerical model simulations suggest WD spin axis inclination, $i$ = 30-50\deg. The broad-band circular polarization spectrum is flatter than predicted by constant temperature cyclotron models, and requires an inhomogeneous emission region with a spread of the physical parameters (temperature, electron density, magnetic field) and possibly a complicated geometry.

Period analysis from timings of the zero-crossover of circular polarization puts strict constraints on the WD spin period changes. The respective spin-down timescales are very long, $T_s > 10^9$ yr, which means that the WD in V405Aur is currently accreting very closely at the spin equilibrium rate. With the spin to orbital period ratio, $P_{spin}/P_{orb} = 0.0365$, such a condition is achieved in disk-type accretion where the angular momentum gain from matter transferred onto the WD via the disk balances the magnetic braking torque. Comparison with numerical accretion model computations (Norton et al. 2008) shows that for V405 Aur equilibrium condition would be obtained with mass ratio $q_1 \sim$ 0.4, if the magnetic field is $B \sim$ 30 MG.

\acknowledgments
The Nordic Optical Telescope is operated on the island of La Palma jointly by
Denmark, Finland, Iceland, Norway, and Sweden, in the Spanish Observatorio
del Roque de los Muchachos (ORM) of the Instituto de Astrophysica de Canarias.
The KVA-60 telescope is operated by Tuorla Observatory of the University of
Turku, at ORM under the agreement between the University of Turku, Finland, and
the Royal Academy of Sciences, Sweden (Kungliga Vetenskapsakademien).
\end{document}